\documentclass[twocolumn,showpacs,prl,superscriptaddress]{revtex4}

\usepackage{graphicx}

\usepackage{amssymb}
\usepackage{amsmath}
\usepackage{amsfonts}
\usepackage{mathrsfs}

\renewcommand{\epsilon}{\varepsilon}

\begin{document}
\title{Driven Diffusion in Periodic Potentials
with Stochastic Path Integral Hyperdynamics}

\author{Mahendra D. Khandkar}
\affiliation{Department of Applied Physics, COMP Center of
Excellence, Helsinki University of Technology, P.O. Box 1100,
FI-02015 TKK, Espoo, Finland}

\author{L.Y. Chen}
\affiliation{Department of Physics, University of Texas at San
Antonio, San Antonio, Texas 78249-0697}

\author{S.C. Ying}
\affiliation{Department of Physics, Box 1843, Brown University,
Providence, Rhode Island 02912-1843}

\author{T. Ala-Nissila}
\affiliation{Department of Applied Physics, COMP Center of
Excellence, Helsinki University of Technology, P.O. Box 1100,
FI-02015 TKK, Espoo, Finland}
\affiliation{Department of Physics,
Box 1843, Brown University, Providence, Rhode Island 02912-1843}

\date{May 29, 2009}
\begin{abstract}

We consider the driven diffusion of Brownian particles in 1D
periodic potentials using the recently proposed Stochastic Path
Integral Hyperdynamics (SPHD) scheme [L.Y. Chen and L.J.M. Horing,
J. Chem. Phys. {\bf 126}, 224103 (2007)]. First, we consider the
case where a single Brownian particle is moving in a spatially
periodic potential and subjected to an external ac driving force. We
confirm that there is no stochastic resonance in this system and
find that at higher frequencies the diffusion coefficient $D$ is
strongly suppressed. The second case is that of a dimer moving in a
periodic potential with a static bias. For this case, there's a
strong suppression of $D$ when the dimer bond length is an integer
multiple of the lattice constant of the potential. For both cases,
we demonstrate how the SPHD allows us to extract the dynamical
information exactly at different bias levels from a single
simulation run, by calculating the corresponding statistical
re-weighting factors.
\end{abstract}


\maketitle

\section{Introduction}

The study of particles performing Brownian motion in a periodic
potential constitutes a hallmark example of stochastic particle
dynamics with important applications in various branches of science
and technology. Perhaps the most common application of periodic
Brownian motion is the diffusive dynamics of atoms and molecules on
crystal surfaces \cite{review}. Surface diffusion is among the most
important mechanisms that controls processes such as island
nucleation and subsequent surface growth. It has been shown that by
controlling the mobility of particles on the surface by external
means, such as an ac or dc electric field, allows morphological
control over the growing surfaces \cite{review}. It is thus of great
interest to model periodic Brownian motion with static and
time-dependent external fields.

To this end, there have been several studies reporting the diffusion
of a single Brownian particle in a periodic potential with external
ac bias applied \cite{acbias1,acbias2,acbias3,acbias4}. The case of
a dimer consisting of two connected particles has also been
considered \cite{dimer1,dimer2,dimer3}, in which case there's an
additional length scale in the problem, namely the dimer bond
length, as shown schematically in Fig. 1. Most of the studies
reporting the behavior of Brownian particles discuss the influence
of an oscillating bias on transport coefficients. The central issue
here is existence of a stochastic resonance (SR), which can greatly
enhance the diffusion coefficient $D$ in 2D \cite{acbias2}. However,
it has been shown in the case of 1D periodic potentials that
although the local jump rate of particles can be enhanced, there is
no true SR in the hydrodynamic limit \cite{acbias1,acbias4}.
\begin{figure}
\includegraphics[height=5cm, width=7cm]{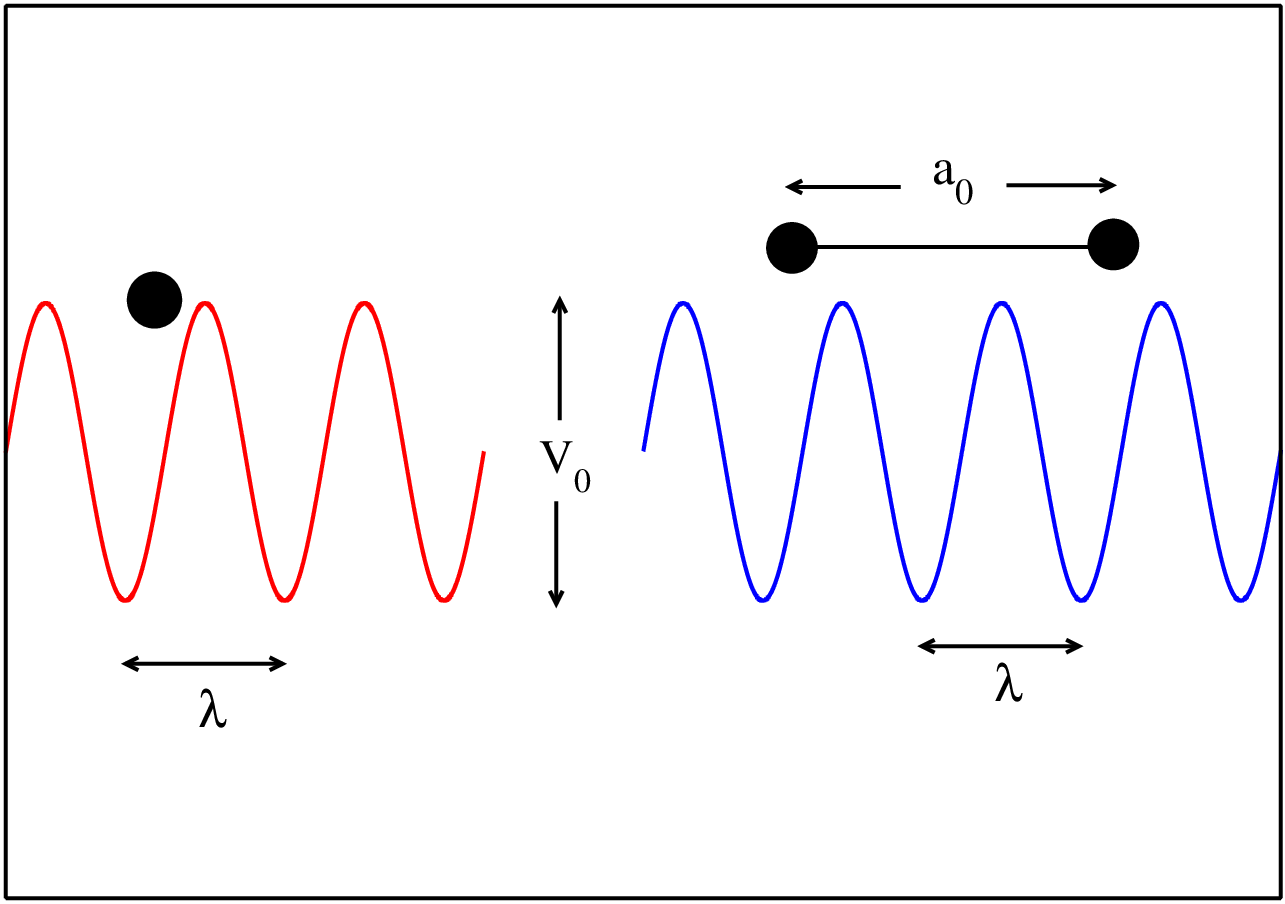}
\caption{Driven diffusion in a 1D periodic potential. The relevant
parameters are the barrier height $V_0$, the lattice constant
$\lambda$ and the dimer bond length $a_0$. The details of an
external bias are discussed in the text.}
\end{figure}
An interesting limit of the periodic Brownian motion is where the
energy barrier $V_0$ is much larger than the thermal energy, {\it
i.e.} $\beta V_0 \gg 1$ \cite{review}, where $\beta=1/k_BT$ and
$k_B$ is the Boltzmann constant and $T$ the temperature (cf. Fig.
1). Since Brownian motion is activated by thermal fluctuations, the
diffusion rate is proportional to $\exp(-\beta V_0)$ which becomes
very small at low temperatures. To overcome this rare event problem
in Molecular Dynamics (MD) simulations, Voter \cite{voter} has
proposed the so-called Hyperdynamics (HD) scheme, which involves
accelerating the dynamics by adding proper bias potential, which
effectively lowers the barrier height. The dynamics is then
corrected based on the approximate Transition State Theory (TST).
There exist various approaches to the choice of the bias potential,
and some examples can be found in Refs.
\cite{hyper1,hyper2,hyper3,hyper4}.

However, recently a new scheme has been proposed that is based on
the mapping of the stochastic Langevin equation to a path integral
form \cite{SPHD}. Unlike the standard HD scheme, this so-called
Stochastic Path Integral Hyperdynamics (SPHD) method allows an
\emph{exact} correction of the dynamics by resampling the simulated
paths. In other words, this method is not restricted to the TST
approximation. Further, it is not restricted to static energy
barriers; both entropic barriers and even time-dependent bias can be
employed. This allows an efficient way to overcome the large barrier
problem, as demonstrated in Ref. \cite{SPHD}. However, the SPHD
method is not limited to the case of high barriers. Since in
principle any external bias force can be used, it should be possible
to obtain results for \emph{many different bias values} from running
LD simulations with a single value of the bias force, or even
without such a force if need be. To demonstrate this, in this work
we have undertaken to employ the SPHD method to study periodic
Brownian motion for two interesting cases. In the first case, we
consider the diffusion of a single Brownian particle in a 1D
periodic potential with ac forcing. The second case is that of a
dimer diffusing in a 1D periodic potential with a static bias. For
both cases, we show how the SPHD method can be efficiently employed
to obtain the transport coefficients for a range of different
external forcing terms from single simulation runs.

\section{Stochastic Path Integral Hyperdynamics}

Brownian motion of a single particle can be represented by the
Langevin equation
\begin{equation}
m \ddot{r}(t)+m \gamma \dot{r}(t)-F=\xi (t),
\end{equation}
where $r(t)$ denotes the position of the particle (of unit mass m)
at time t, moving under the influence of external force $F$ and the random
noise term $\xi(t)$ satisfies $\left\langle \xi (t)\right\rangle =0$
and $\left\langle \xi (t)\xi (t')\right\rangle =2k_{B}T m \gamma \delta
(t-t')$, where $\gamma$ denotes the friction coefficient. In
general, to accelerate the dynamics one must add a bias force
$F_{b}(r,t)$ to the Langevin equation to get
\begin{equation}
m \ddot{r}(t)+m \gamma \dot{r}(t)-F-F_{b}(r,t)=\xi(t).
\end{equation}
With an appropriately chosen bias potential, the dynamics of the
system evolves much faster than in the original Langevin equation.
The dynamics given by numerically solving Eq. (2) is fictitious, of
course. However, the use of path integral formalism allows an exact
compensation of the effect of adding the bias force by defining an
effective action functional $I_{\xi}(t)$ \cite{SPHD}
\begin{equation}
I_{\xi}(t)=\frac{1}{4\gamma }\sum_{i}
-F_{b}(r,t_{i})\left[-F_{b}(r,t_{i})-2\xi (t_{i})\right]\Delta t.
\end{equation}
This expression constitutes time integration for a given realization
of the random noise force $\xi$ along a given trajectory $r(t)$. To
recover true dynamics in the absence of $F_{b}(r,t)$, one has to
estimate the SPHD statistical weight factor $\exp(-\beta I_{\xi})$
and simply use it to re-weight every sampled trajectory.

The fundamental quantity associated with Brownian dynamics is the
single-particle (tracer) diffusion coefficient \cite{review}, which
can be defined through the mean square displacement (MSD) of the
tracer particle as
\begin{equation}
 D=\lim_{t \rightarrow \infty} \frac{1}{2t}
  \langle \left[r(t)-r(0)\right]^{2} \rangle.
\end{equation}
When studying particle diffusion the mean square displacement (MSD)
at zero bias (true dynamics) can be obtained by running SPHD with a
bias, calculating $I_{\xi}(t)$ along every trajectory and
re-weighting as
\begin{equation}
\left\langle \left[r(0)-r(t)\right]_{0}^{2}\right\rangle
=\left\langle \left[r(0)-r(t)\right]_{\mathit{HD}}^{2}e^{-\beta
I_{\xi }}\right\rangle,
\end{equation}
where the subscripts $0$ and {\it HD} correspond to the quantities
with zero bias (true dynamics) and finite bias, respectively. The
SPHD has an additional powerful feature which can be seen from the
biased Langevin equation and the expression for the re-weighting
factor $I_{\xi}(t)$. It is possible to obtain the true dynamics for
\emph{any} value of the bias force $0 \le f_b \le F_b$ by estimating
$I_{\xi}(t, f_{b})$ corresponding to some $f_{b}$. With this, the
true dynamics at multiple bias levels ($f_{b}$) can be extracted,
 simultaneously, from a single simulation run of
 biased dynamics according to Eq. (2) and estimating the
weight factors, corresponding to multiple values of $f_{b}$ along all
the trajectories.

\section{Results and Discussion}

\subsection{Brownian Particle in Periodic Potential with
Time-Varying Bias}

The first case where we consider the application of the SPHD method
is that of a Brownian particle in a one-dimensional spatially
periodic potential with an external, time-dependent ac driving force
\cite{acbias1,acbias2,acbias3,acbias4}. For such a system, the
equation of motion is given by
\begin{equation}
m \ddot{x}(t)+m \gamma \dot{x}(t)-F=\xi (t)+ A\sin (2\pi \omega t),
\end{equation}
where the second term on the right hand side indicates an ac driving
force with amplitude $A$ and frequency $\omega$. The diffusion of a
Brownian particle can be studied with respect to various values of
these two parameters. $F$ is force due to spatially periodic
potential $-(V_0/2)[1-\cos(2 \pi x/\lambda) ]$. Here, we have
employed the SPHD method by numerically solving Eq. (6) with $A = 0$
using the standard velocity Verlet scheme \cite{allen}. The
diffusion coefficients for different values of $A$ and $\omega$ can
be obtained by choosing $f_b(A,\omega) = A \sin (2 \pi \omega t)$
and then estimating the correction factor $I_\xi(t,f_{b})$ and the
reweighing factor $\exp(-\beta I_\xi )$ for every bias force.

We set scales for length = $\lambda$, energy = $k_BT$ and mass = $m$.
Then time unit is defined as $t = \lambda \sqrt{(m /k_BT)}$
and all other relevant quantities are expressed as dimensionless and 
indicated by a tilde over the respective symbols. 

The parameters we have used in the present work are $\tilde{V_0}=2$, 
 and $\tilde{\gamma} = 2$. The time step we have chosen
is $\Delta \tilde{t} = 0.0005$ \cite{timestep}. From the unbiased runs (in
this case with $\tilde{A}=0$) we have computed the MSD and the
corresponding diffusion coefficients as given in Eq. (4) for a
range of values of $\tilde{A}$ ($\tilde{A} = 0.1, 0.5, 1.0, 1.5$) and the
frequency, as summarized in Fig. 2. All the data have been
averaged over $10^6$ trajectories.
\begin{figure}
\includegraphics[height=5cm, width=7cm]{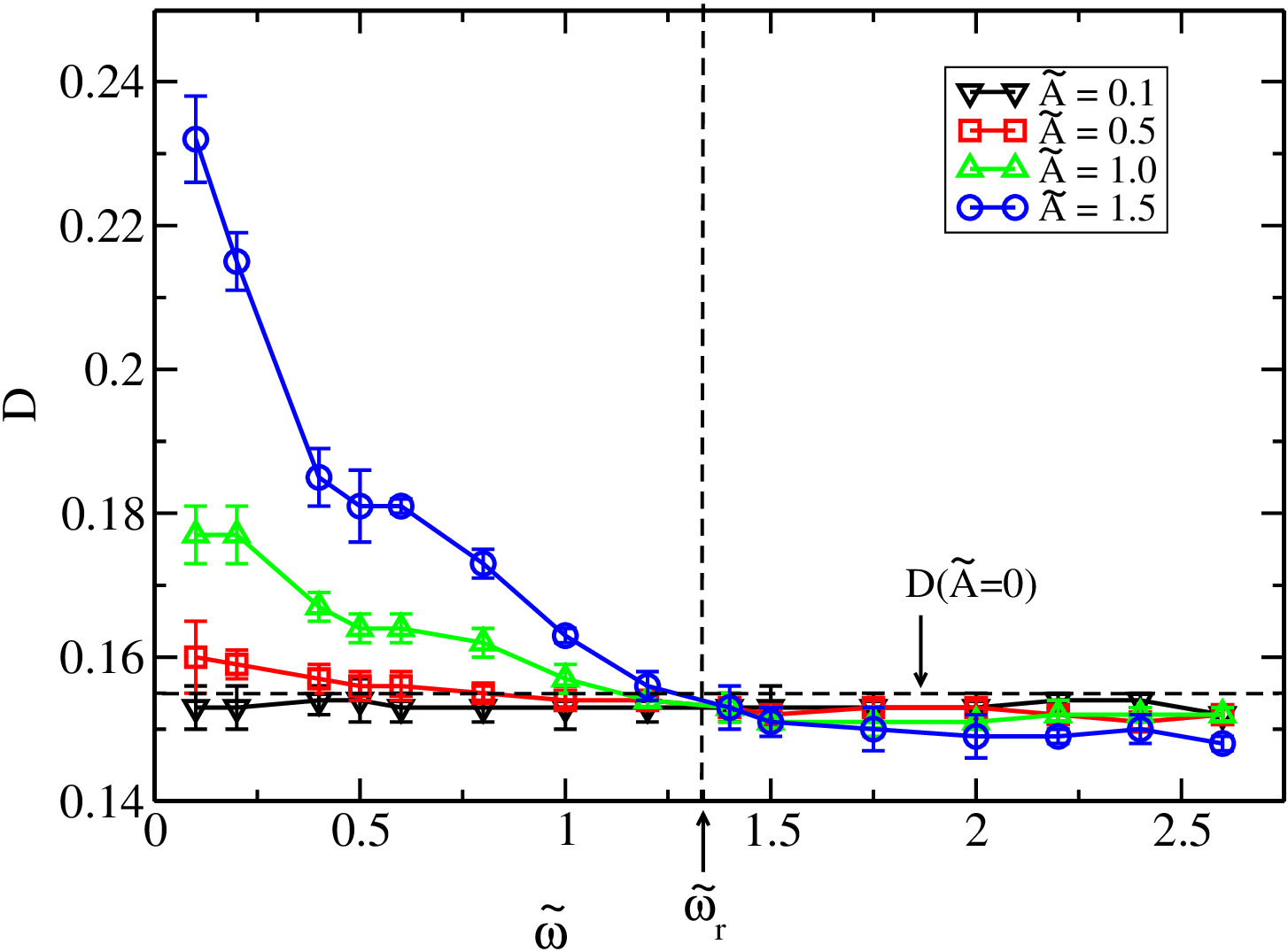}
\caption{Diffusion coefficients for the ac-driven Brownian
particle in a periodic potential as a function of the driving
frequency $\tilde{\omega}$ for various values of the amplitudes $\tilde{A}$. The
horizontal dashed line shows the reference value of $D$ for $\tilde{A}=0$,
and the vertical dashed line indicates the position of the
resonance frequency $\tilde{\omega_r}$ (see text for details).
The solid lines are just guide to an eye and contain no other significance.}
\end{figure}
Our results demonstrate that for a given amplitude $\tilde{A}$ of the
external ac driving force, $D$ decreases monotonically with the
frequency $\tilde{\omega}$. For higher values of $\tilde{\omega}$ studied, it
appears that large values of amplitude $\tilde{A}$ lead to a lower $D$ than
small values of $\tilde{A}$. It seems that the ac bias for higher values of
both $\tilde{A}$ and $\tilde{\omega}$ acts detrimental to activation in the
diffusive motion. This feature can be effectively used to localize
the motion of the particle.

It is of interest to compare the SPHD with direct simulations for
the Langevin equation. In this case, a larger time step of $\Delta
\tilde{t} = 0.005$ was sufficient, and averages were taken over $10^6$
trajectories. In Table I we show results for the case of $\tilde{\omega} =
0.2$ as a function of increasing amplitude $\tilde{A}$. We find excellent
agreement between these data. For the highest value of $\tilde{A}$,
studied here, it appears that SPHD gives a slightly smaller
estimate of $D$ than the LE, although the data still agree within
the error bars. Our preliminary results indicate that with
increasing bias path sampling should be increased, too, which is
caused by the exponential decrease in the reweighing factor with
increasing $\tilde{A}$.
\begin{center}
\begin{table}
\caption{Values of the diffusion coefficient $D$ from SPHD and
direct solution of Eq. (6) with $\tilde{\omega}=0.2$.}
\begin{tabular} {|c c| c |c c|}
\hline
$\tilde{A}$  & &  $D$  & & $D$  \\
          & &   (from Eq. (6)) & &  (SPHD)  \\
\hline
 0.1 & & $0.155 \pm  0.001$ & &  $0.153 \pm 0.002$  \\
\hline
 0.5 & & $0.163 \pm 0.002$ & &  $0.159 \pm 0.004$  \\
\hline
 1.0 & & $0.185 \pm 0.003$ & &  $0.178 \pm 0.005$   \\
\hline
 1.5 & & $0.225 \pm 0.003$ & &  $0.215 \pm 0.007$  \\

\hline
\end{tabular}
\end{table}
\end{center}
An interesting issue in Brownian motion under time-periodic
forcing concerns the existence of stochastic resonance, which
leads to a significant enhancement of the relevant transition
rates \cite{acbias4}. In the case of a double-well potential, SR
is expected to occur in the vicinity of the matching condition
$\omega_r = \pi r_e$, where $r_e$ is the (thermal) escape rate
\cite{acbias4}. In the case of an extended periodic potential
there's enhanced escape for local diffusion jumps over the barrier
$V_0$ \cite{acbias1}. However, it has been shown in Refs.
\cite{acbias1,acbias2,acbias4} that this enhancement exactly
cancels out in the hydrodynamic limit for a 1D periodic potential
such as used in the present study. One can estimate the
frequency ($\omega_{r}$) as \cite{acbias4}
\begin{equation}
\tilde{\omega_r} = (\frac{\pi^2 \tilde{V_0}}{\tilde{\gamma}}) e^{-\tilde{V_0}}
\end{equation}
The calculation results in $\tilde{\omega_{r}} = 1.336$. As mentioned above,
there is no peak in the value of $D$ around $\tilde{\omega_{r}}$ which
confirms the absence of stochastic resonance.

\subsection{Brownian Dimer in Tilted Periodic Potentials}

The second case that we consider here is that of a Brownian dimer
diffusing along a tilted 1D periodic potential (cf. Fig. 1). The
equations of motion for the beads are given by
\begin{equation}
m \ddot{x}_i(t)+ m \gamma \dot{x}_i(t) - F =\xi_i(t) - \nabla_{x_i}
[\frac{V_0}{2}(1-\cos(2 \pi x_i/\lambda)) - b_{t}x_i],
\end{equation}
where $i=1,2$ and the variables $x_i(t)$ denote the positions of the
two beads. The second term on right hand side denotes the force
associated with external static bias with spatial periodicity
$\lambda$ (see Fig. 1) and tilt $b_{t}$. On the left hand side, $F$
represents the force due to the interaction between the constituent
monomers of the dimer which is given by a combination of the
Lennard-Jones (LJ) potential and the FENE (Finitely Extensible
Non-linear Elastic) potential. The LJ potential is given by
\begin{eqnarray}
U_{\mathit{LJ}}(r) & =& 4\epsilon \left[\left(\frac{\sigma
}{r}\right)^{12}-\left(\frac{\sigma }{r}\right)^{6}\right]+\epsilon,
{\rm for} \ r\le 2^{1/6}\sigma; \nonumber \\
      & = & 0, {\rm \ \ otherwise}.
\end{eqnarray}
Here, $\epsilon$ and $\sigma$ define the energy and length scales,
respectively, and $r$ is the separation between the monomers. The
FENE potential is given by
\begin{equation}
U_{\mathit{FENE}}(r)= - \frac{1}{2}kR_{0}^{2}\ln
(1-r^{2}/R_{0}^{2}),
\end{equation}
where $k$ is the effective spring constant and $R_{0}$ sets the
maximum allowed separation between the monomers.

The SPHD method was employed by numerically integrating Eq. (8)
corresponding to a spatially periodic surface with given $\lambda$
and zero tilt.  The dynamics along tilted periodic potentials was
recovered by setting $f_{b}(b_{t}) = -\nabla_{x_i} [- b_{t}x_i]$.
For the resampling, the corresponding weight factors were evaluated
as in the previous section.

Again, we set scales for length = $\sigma$, energy = $\epsilon$ and mass = $m$.
Then time unit is defined as $t = \sigma \sqrt{(m /\epsilon)}$
and all other relevant quantities are expressed as dimensionless and 
indicated by a tilde over the respective symbols. 

The parameters were set to be $\tilde{V_0} = \tilde{T} = 0.1$, $\tilde{\gamma} = 1$ and
the time step $\Delta \tilde{t} = 0.0005$, with averages taken over $10^6$
trajectories. 
The FENE parameters are $\tilde{k}=1$ and $\tilde{R_{0}} = 2$. The periodicity of 
the 1D potential was changed between $\tilde{\lambda} = 1/2$ and $2$, and 
the tilt parameters were $\tilde{b_t}= 0.0, 0.05, 0.1$. The effective diffusion coefficient was
estimated by using the relation
\begin{equation}
 D=\lim_{t \rightarrow \infty} \frac{\left\langle
x_{\rm cm}^{2}(t)\right\rangle -\left\langle x_{\rm
cm}(t)\right\rangle^{2}}{2t},
\end{equation}
where $x_{\rm cm}$ is the center of mass of the dimer, in order to
subtract the drift term caused by $\tilde{b_t} > 0$.
\begin{figure}
\includegraphics[height=5cm,width=7cm]{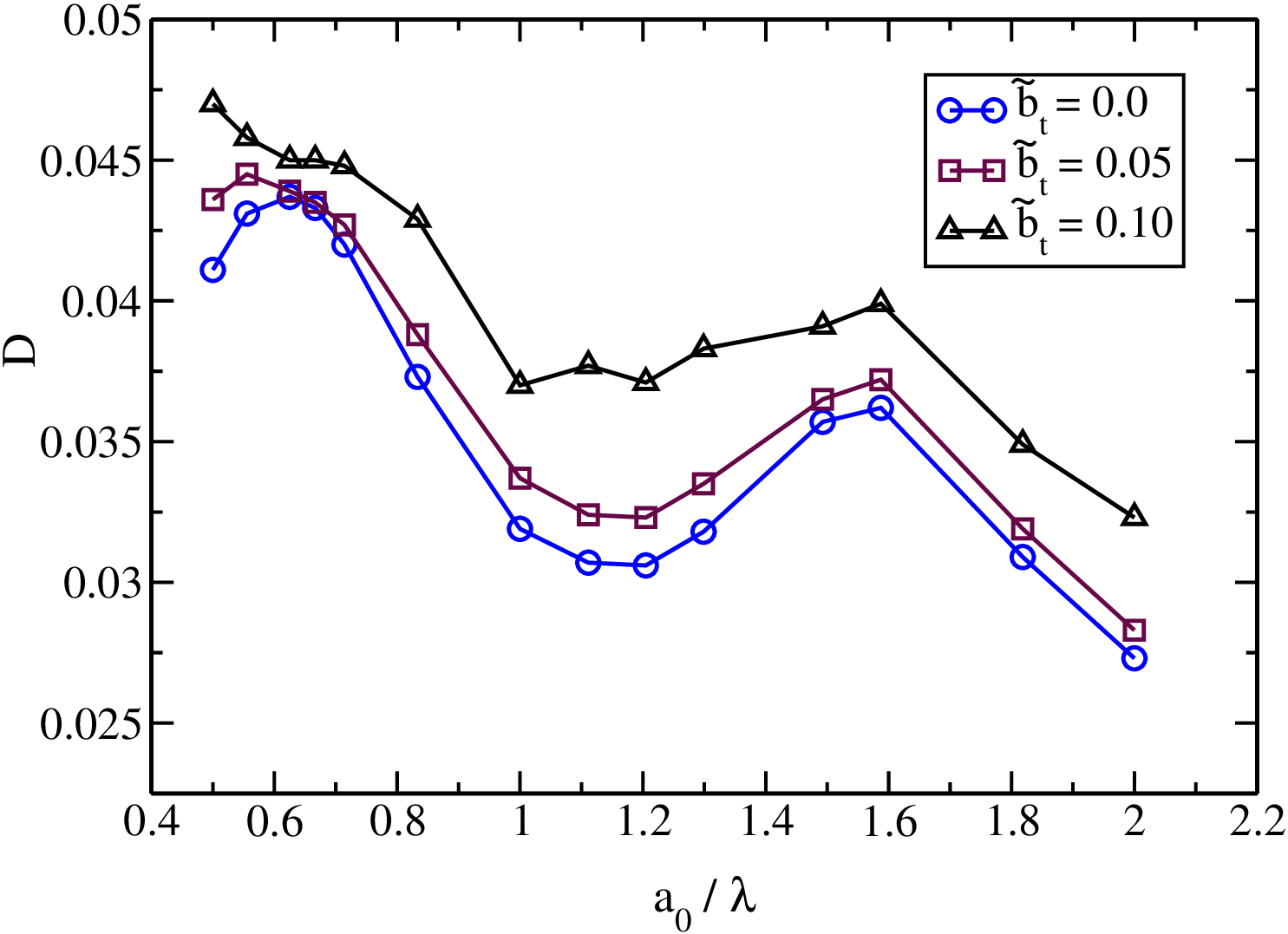}
\caption{The dimer diffusion coefficient $D$ \emph{vs.} the
wavelength $a_0/\lambda$ for various tilts. Error bars are of the
order of the size of the symbols, or smaller.
The solid lines are just guide to an eye and contain no other significance.}
\end{figure}
In Fig. 3 we summarize our results for $D$ as a function of the
wavelength of the periodic potential for three tilt values. For the
case of a dimer, there are now two relevant length scales in the
system (see Fig. 1): the zero temperature equilibrium bond length of the dimer 
$\tilde{a_0} \approx 1.10$ and the wavelength $\tilde{\lambda}$ of the underlying
periodic potential.
Thus, for the dimer motion
there's a matching of the two lengths when the ratio $a_0/\lambda=n$
is an integer. On the other hand, when this ratio is a half-integer,
there's strong competition between the dimer bond and potential
energy. In the latter case, it should be easier for the dimer to
escape as the effective diffusion barrier is lower. Indeed, as seen
in Fig. 3 we find that for every value of the tilt there is a strong
decrease in $D$ in the vicinity of the first matching condition
$a_0/\lambda=1$ \cite{temperature}. On the other hand, near
$a_0/\lambda = 3/2$ there is a local maximum in $D$, as expected,
followed by another pronounced minimum near $a_0/\lambda = 2$. In our model,
the interaction between the monomers is highly anharmonic. This results in an equilibrium
 dimer separation which is temperature dependent and the first minimum in D in Fig.3 is 
shifted from the zero temperature value $a_0/\lambda = 1$ to 
$a_0/\lambda = 1.2$. A finite tilt increases the overall magnitude of
$D$ while its non-monotonic behavior as a function of $a_0/\lambda$
prevails almost unchanged .
Our results are consistent with the Langevin dynamics studies of Bammert
\emph{et al.} \cite{dimer1}, who considered dimer diffusion in a 2D
periodic square potential (with a hydrodynamic interaction term
included) and found that there's a local maximum in $D$ for
$a_0/\lambda = 3/2$. Heinsalu \emph{et al.} \cite{dimer2} have
reported dimer diffusion on a 1D washboard-like potential, and they
also find a flat minimum at $a_0/\lambda=1$ for small tilts.

Another interesting feature in our results is that the minima in $D$
deepen with increasing $n$. This can be understood as follows. When
the matching condition is met (\emph{i.e.} $n$ is an integer), both
beads in the dimer can sit exactly at the minima of the potential
separated by $n-1$ minima. The elementary diffusion move of the
dimer (local jump rate) consists of both of the beads crossing the
saddle points (potential maxima) synchronously (assuming that bond
fluctuations can be neglected). For every $n$ the effective
diffusion barrier is exactly the same independent of $n$. However,
the jump length of the dimer depends on $n$ (in units of $\tilde{a_0}$) such
that the distance the dimer moves is given by $\tilde{a_0}/n$. Thus, if we
use the Dynamical Mean Field theory to approximate the diffusion
coefficient \cite{jumps}, the prefactor of $D$ is proportional to
the jump length squared, which gives
\begin{equation}
D \propto \frac{1}{n^2}.
\end{equation}
We have analysed the data upto around  $a_0/\lambda = 4$ and checked that 
minima of $D$ near integer values of $a_0/\lambda$ indeed decreases with 
increasing $n$, but slower than predicted by Eq. (12) for the present set of parameters.
Indeed, Eq. (12) strictly holds only in the limit of a rigid dimer
bond and $\tilde{V_0} \gg 1$, which is not in the range of the
parameters used here.

%
%

In analogy to the single particle case, we have compared our results
from the SPHD scheme to data obtained from directly integrating the
Langevin equation with the static bias for the largest value of the
tilt $\tilde{b_t}= 0.10$, for various values of $\tilde{\lambda}$. For the straight
LD simulations $\Delta \tilde{t} = 0.005$ and averages were taken over
$10^{6}$ trajectories. There is again good agreement between these
two sets of data.

An intriguing further extension of the SPHD method is extrapolation
with more than one bias force parameter. In the present case, we did
some test runs by numerically integrating Eq. (8) corresponding to a
flat surface \emph{and} zero static bias ($V_0=b_t=0$). The dynamics
along tilted periodic potentials was recovered by setting
$f_{b}(\lambda, b_{t}) = -\nabla_{x_i}
 [(V_0/2)(1-\cos(2 \pi x_i/\lambda)) - b_{t}x_i]$, and we used the
same parameters as in the single-parameter resampling case. We found
that using $10^6$ trajectories the discrepancies with respect to the
data in Table II were about 16 \% at largest. This is due to the
fact that when simulations are run on a smooth surface, all
trajectories of the dimer are weighted equally. However, in the
actual periodic potential the main contribution to the diffusion
coefficient comes from paths crossing the saddle point from one
minimum to another. Thus, in the resampling procedure most of the
paths are not relevant for determining the value of $D$, and thus
the errors remain relatively large even with $10^6$ paths for the
present case.

\begin{center}
\begin{table}
\caption{Comparison between results for the diffusion coefficient
$D$ obtained from SPHD and from Eq. (8) with tilt $\tilde{b_t} = 0.10$.}
\begin{tabular} {|c c | c | c c|}
\hline
$a_0/\lambda$  & &  $D$  & & $D$  \\
          & &   (Langevin eq.)  & & (SPHD)  \\

\hline

 1/2 & &  $ 0.0482 \pm 0.0004$   & &  $ 0.0470 \pm 0.0005$ \\
\hline
 2/3 & &  $ 0.0454 \pm 0.0005$   & &  $ 0.0450 \pm 0.0004$ \\
\hline
 1 & &  $ 0.0368 \pm 0.0003$   & &  $ 0.0370 \pm 0.0004$ \\

\hline
 3/2  & &  $ 0.0393 \pm 0.0004$   & &   $ 0.0391\pm 0.0006$ \\

\hline
 2 & &  $ 0.0320 \pm 0.0004$   & &  $ 0.0323 \pm 0.0006$ \\

\hline
\end{tabular}
\end{table}
\end{center}
%

%
%
%
%
%
%
%
%
%
%

\section{Summary and Conclusions}

In this work, we have employed the recently proposed SPHD scheme to
study the diffusive motion of Brownian particles in periodic
potentials in 1D. Unlike the HD schemes proposed so far, the SPHD
scheme allows an exact correction of the biased dynamics based on
reweighing of all the transition paths. There's also no restriction
for the type of bias potential used, and thus the SPHD method can be
used to extrapolate results to \emph{multiple} values of the bias
force from a single simulation run, as shown by the two cases we
have studied here: time-dependent forcing for a single Brownian
particle and constant forcing for a Brownian dimer. For a Brownian
particle in an external ac bias, our results are in agreement with
previous studies and show that there's no stochastic resonance in
this system. For the second case of a dimer moving in a periodic
potential with a static bias we find a strong suppression of $D$
when the dimer bond length is an integer multiple of the lattice
constant of the potential. This suppression is weakened by an
applied dc bias. Our work demonstrates how external forcing can be
used to control particle mobilities in periodic potentials.

\section{Acknowledgements}

We wish to thank Kaifu Luo and Jaeoh Shin for their helpful
suggestions. This work has been supported in part by The Academy of
Finland through its Centre of Excellence (COMP) and TransPoly
Consortium grants. We also thank CSC-The Centre for Scientific
Computing Ltd. for allocation of computational resources.


%

\begin{thebibliography}{99}

\bibitem{review} T. Ala-Nissila, R. Ferrando and S.C. Ying, Adv. Phys. \textbf{51},
\ 949 (2002).

\bibitem{acbias1} Kallunki, J., Dub\'e, M., and Ala-Nissila, T.,
Surf. Sci. \textbf{460}, 39 (2000).

\bibitem{acbias2} Zhang and Bao, Surf. Sci. \textbf{540}, 145
(2003).

\bibitem{acbias3} L.Y. Chen and P.L. Nash, J. Chem. Phys. \textbf{121}, 3984
(2004).

\bibitem{acbias4} J. Kallunki, M. Dub\'e, and T. Ala-Nissila, J. Phys.: Cond.
Mat. \textbf{11}, 9841 (1999).

\bibitem {dimer1} J. Bammert, S. Schreiber and W. Zimmermann, Phys. Rev. E
\textbf{77}, 042102 (2008).

\bibitem{dimer2} E. Heinsalu, M. Patriarca and F. Marchensoni, Phys. Rev. E
\textbf{77}, 021129 (2008).

\bibitem{dimer3}  O.M. Braun, Phys. Rev. E \textbf{68}, 051101
(2003).

\bibitem{voter} A.F. Voter, J. Chem. Phys. \textbf{106}, 4665
(1997).

\bibitem{SPHD} L.Y. Chen and N.J.M. Horing, J. Chem. Phys.
\textbf{126}, 224103 (2007).

\bibitem{hyper1}\ D. Hammelberg, J. Mongan and J.A. McCammon, J. Chem. Phys.
\textbf{120}, 11919 (2004).

\bibitem{hyper2} J.C. Wang, S. Pal and K.A. Fichthorn, Phys. Rev. B \textbf{63},
085403 (2001).

\bibitem{hyper3} J.A. Rahman and J.C. Tully, J. Chem. Phys. \textbf{116}, 8750
(2002).

\bibitem{hyper4} R.I. Cuckier and M. Morillo, J. Chem. Phys. \textbf{123},
234908 (2005).

\bibitem{allen} M.P. Allen and D. J. Tildesley,
{\it Computer Simulation of Liquids}, Oxford: Clarendon (1994).

\bibitem{timestep} We have found that for an accurate evaluation of
the re-weighting factor for long enough times a smaller time step
is required here than for direct solution of the Langevin eqution.

\bibitem{temperature} Here $\tilde{V_0} = 1$ and thus the
minimum is extended towards higher values of $a_0/\lambda$ due to
the asymmetry of the dimer bond potential. We have checked that
for $\tilde{V_0} = 5$ the minimum shifts to $a_0/\lambda \approx
1.1$ for $\tilde{b_t}=0$.

\bibitem{jumps} D.A. Reed and G. Ehrlich, Surf. Sci. {\bf 102}, 588
(1981); T. Hjelt, I. Vattulainen, J. Merikoski, T. Ala-Nissila and
S.C. Ying, Surf. Sci. Lett. {\bf 380}, L501 (1997).

\end{thebibliography}
\end{document}